\documentclass[a4paper,11pt]{article}
\pdfoutput=1 % if your are submitting a pdflatex (i.e. if you have
\usepackage{jinstpub} % for details on the use of the package, please
\usepackage{graphicx}
\graphicspath{{fig/}}
\usepackage{siunitx}
\usepackage{hyperref}
\usepackage{slashbox} % for slash splitting table cells
\hypersetup{hidelinks,colorlinks=false,breaklinks=true,urlcolor=ocre}

\title{\boldmath Investigation of Amorphous Germanium Contact Properties with Planar Detectors Made from USD-Grown Germanium Crystals}

\author[a] {W.-Z. Wei}
\author[a] {, X.-H. Meng}
\author[a] {, Y.-Y. Li}
\author[a] {, J. Liu}
\author[a] {, G.-J. Wang}
\author[a] {, H. Mei}
\author[a] {, G. Yang}
\author[a,b,1]{, D.-M. Mei\note{Corresponding author.}}
\author[c] {and C. Zhang} 
\affiliation[a]{Department of Physics, The University of South Dakota,\\414 E. Clark Street, Vermillion, South Dakota 57069, USA}
\affiliation[b]{School of Physics and Optoelectronic, Yangtze University, \\ 1 Nanhuan Street, Jingzhou 434023, China}
\affiliation[c]{Physics, Mount Marty College, \\ 1105 W. 8th St, Yankton, South Dakota 57078, USA}
\emailAdd{Dongming.Mei@usd.edu}

\abstract{The characterization of detectors fabricated from home-grown crystals is the most direct way to study crystal properties. We fabricated planar detectors from high-purity germanium (HPGe) crystals grown at the University of South Dakota (USD). In the fabrication process, a HPGe crystal slice cut from a USD-grown crystal was coated with a high resistivity thin film of amorphous Ge (a-Ge) followed by depositing a thin layer of aluminum on top of the a-Ge film to define the physical area of the contacts. We investigated the detector performance including the $I$-$V$ characteristics, $C$-$V$ characteristics and spectroscopy measurements for a few detectors. The results document the good quality of the USD-grown crystals and electrical contacts.}


\begin{document}
\maketitle
\flushbottom

\section{Introduction}
\label{sec:intro}
 Cosmogenic activation of germanium (Ge) isotopes can limit the sensitivity of ton-scale Ge-based dark matter and neutrinoless double-beta decay experiments~\cite{wei}. Growing Ge crystals and fabricating detectors underground where the experiment will be built is the best option to reduce the background events from cosmogenic activation processes. High-purity Ge (HPGe) crystals have been grown at the University of South Dakota (USD) to pave the way for fabricating Ge crystals and detectors underground~\cite{wang1, wang2, wang3, wang4, wang5}. Low background Ge detectors are a well-accepted methodology in the search for dark matter~\cite{cdms, edeweiss, cdex} and neutrinoless double-beta decay~\cite{klapdor, gerda, mjd}. 
 
 When HPGe detectors are cooled down to liquid nitrogen temperature, they are usually operated in a full depletion mode. The electrical contacts on the detector are required to effectively block charge carrier injection at a sufficiently high voltage to maintain a low-leakage current and have good charge collection. Traditionally, electric contacts on Ge detectors consist of n$^{+}$ contacts formed by lithium diffusion for hole blocking and p$^{+}$ contacts formed by boron ion implantation for electron blocking. This contact technology has been the industry standard since the early days of HPGe detector development due to relatively robust fabrication process. HPGe detectors with these robust contacts have been used widely for applications requiring excellent energy resolution as a spectrometer. However, in addition to measuring the energies of particles produced in rare or novel reactions, there are increasing demands on position-sensitive HPGe detectors for applications requiring particle tracking or imaging in many fields such as nuclear physics, astrophysics, nuclear nonprolifieration and medical imaging. In order to achieve good position resolution, both hole and electron blocking contacts are normally segmented~\cite{del, bog, bur,zio}. Unfortunately, traditional contacts especially the thick lithium-diffused contacts have drawbacks for fine segmentation. Lithium atoms in Ge have high mobility, leading to thicker and wider lithium-diffused contacts with time at room temperature. This makes the fine segmentation of lithium contacts difficult to create and maintain. Although there exists thin layer n$^{+}$ contacts using phosphorus ion implantation~\cite{ben}, the fabrication processes are elaborate and these contacts cannot withstand high electric fields. 
 
 Amorphous Ge (a-Ge) contacts can be an alternative to the traditional contacts discussed above since a-Ge can also provide rectifying contacts on Ge detectors. More importantly, a-Ge deposited on a crystalline Ge surface showed good bipolar blocking behavior, i.e. a-Ge contacts can block both hole and electron injection well. In addition, it is relatively simple to fabricate HPGe detectors with fine segmentation features using a-Ge technology. 
 
 The first experimental investigation of the a-Ge contacts on crystalline Ge was conducted in 1964~\cite{gri}. In 1971, England and Hammer~\cite{eng} first reported the results of successfully using a-Ge as blocking contacts on silicon radiation detectors. Hansen and Hall~\cite{han} investigated the bipolar blocking property of a-Ge contacts on HPGe detectors in 1977. However, their testing results showed large variations in leakage current. In 1992, Luke et al.~\cite{luke1} reported the performance of planar Ge detectors with a-Ge contacts produced using RF sputtering techniques. Based on their preliminary tests, a-Ge contacts on planar Ge detectors exhibit excellent bipolar blocking features. More studies from Luke ~\cite{luke2, luke3} and Amman~\cite{amman1}, published in 1994 and 2000, show that a-Ge contacts can be used to create position-sensitive Ge detectors for x-ray and gamma-ray imaging. Since then, HPGe detectors with a-Ge contacts have been used or are being developed for various applications such as field shaping~\cite{amman2} and proximity electrode signal readout~\cite{luke4, amman3}. More recently, Looker et al. and Amman~\cite{looker,amman5} conducted a systematic study of leakage current in HPGe detectors with amorphous semiconductor contacts using RF sputtering techniques. According to their study, the detector performance, particularly leakage current, is dependent on the fabrication process and operational parameters including the sputter gas pressure and composition, and time spent at room temperature.
 
 In this paper, we report on the performance of HPGe detectors with RF-sputtered a-Ge contacts using the Ge crystal grown by ourselves at USD. The Ge detector testing setup is presented in section~\ref{sec:setup}, followed by the electrical measurements including the $I$-$V$ and $C$-$V$ measurements in section~\ref{sec:elec_mea}. The spectroscopy measurements are presented in section~\ref{sec:spec_mea}. Finally, the conclusions are summarized in section~\ref{sec:conl}.
 
 \section{Detector Characterization Setup}
 \label{sec:setup}
\begin{figure} [htbp]
  \centering
  \includegraphics[width=0.45\linewidth]{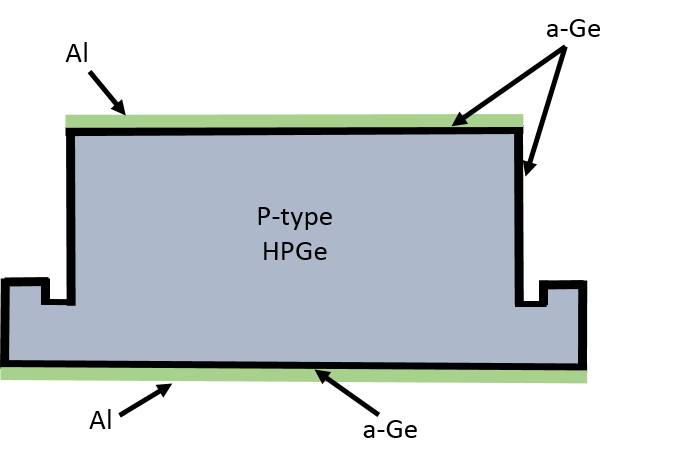}
  \includegraphics[width=0.3\linewidth]{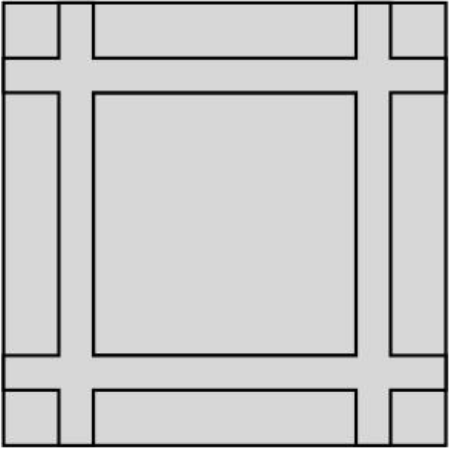}
  \caption{Schematic cross-sectional (left) and top view (right) drawing showing the typical geometry of the HPGe detectors fabricated at USD.}
  \label{fig:geo}
\end{figure}

 We have fabricated a handful of small planar Ge detectors with crystals we grew to examine the fabrication process parameters and the detector performance, so that we will have a better understanding of the fabrication process and particularly the properties of the crystals we grew. All of these materials were p-type with approximate impurity concentrations from $5\times$10$^{9}$ to $5\times$10$^{10}$/cm$^3$. Several crystals slices were cut from three different boules. The main body of each detector crystal consisted of a rectangular contact area measuring from $\sim$ 0.8 to 1.8 cm on a side and a thickness perpendicular to the contact faces of about 0.5 to 1.5 cm (see figure~\ref{fig:geo}). 
%\begin{figure} [htbp]
  %\centering
  %\includegraphics[clip,width=0.5\linewidth]{geometry_1.png}
  %\caption{Schematic cross-sectional drawing showing the typical geometry of the HPGe detectors fabricated at USD.}
 % \label{fig:geo}
%\end{figure}

In order to convert an HPGe crystal into a planar detector, the following procedures were followed: (1) the crystal was cut to the desired shape as shown in figure~\ref{fig:geo} utilizing a diamond saw; (2) each exposed surface of the cut crystal was lapped to remove any blade marks from cutting; (3) the crystal was etched in a 4:1 nitric to hydrofluoric acid mixture to remove any surface damage introduced by those mechanical processes; (4) if all surfaces looked good, the crystal was then etched briefly in fresh 4:1 etchant, rinsed in deionized water, and blown dry with nitrogen to prepare the surfaces for the electrical contact deposition; (5) the crystal was loaded into an RF diode sputtering system immediately after step 4) for the coating of all surfaces with a-Ge (the top and side surfaces were coated first, and then the crystal was flipped over for the coating of bottom surface); (6) the crystal was loaded into an E-beam machine for Al evaporation onto both the top and bottom surfaces of the crystal. The recipe used for a-Ge deposition for all planar detectors made in our lab was the same and it was the one recommended by Amman, who did a lot of work~\cite{amman5} to find a good recipe that yields the lowest leakage current. This recipe consisted of depositing the a-Ge with a 14 mTorr pressure of Ar with 7\% H$_{2}$ gas mixture at a power of 100 W. The thickness of the a-Ge film produced by this recipe is about 300 nm. Please refer to our work~\cite{lily} for the fabrication details for all detectors fabricated in our lab.

\begin{figure} [htbp]
  \centering
  \includegraphics[clip,width=0.6\linewidth]{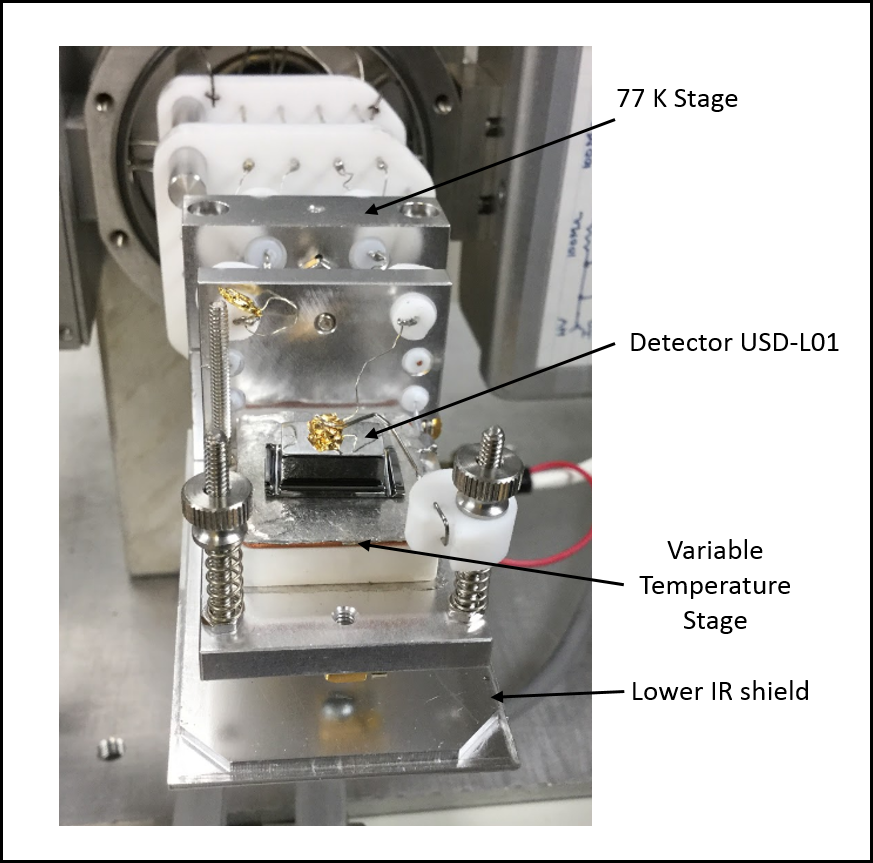}
  \caption{Detector USD-L01 was loaded into the test cryostat.}
  \label{fig:cryo}
\end{figure}

\begin{figure} [htbp]
  \centering
  \includegraphics[clip,width=0.9\linewidth]{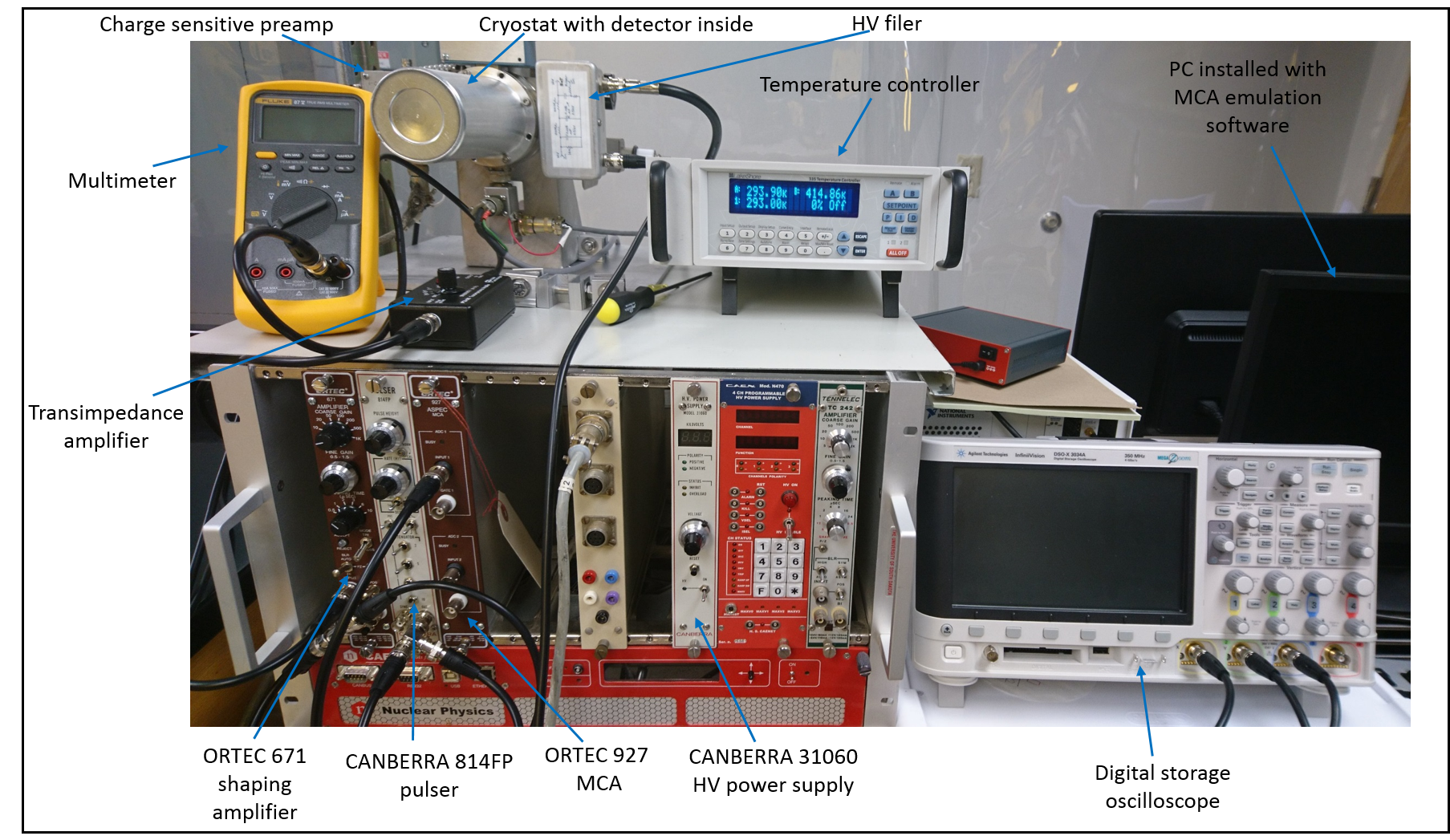}
  \caption{Experimental setup for detector characterization.}
  \label{fig:setup}
\end{figure}

\begin{figure} [htbp]
  \centering
  \includegraphics[clip,width=0.9\linewidth]{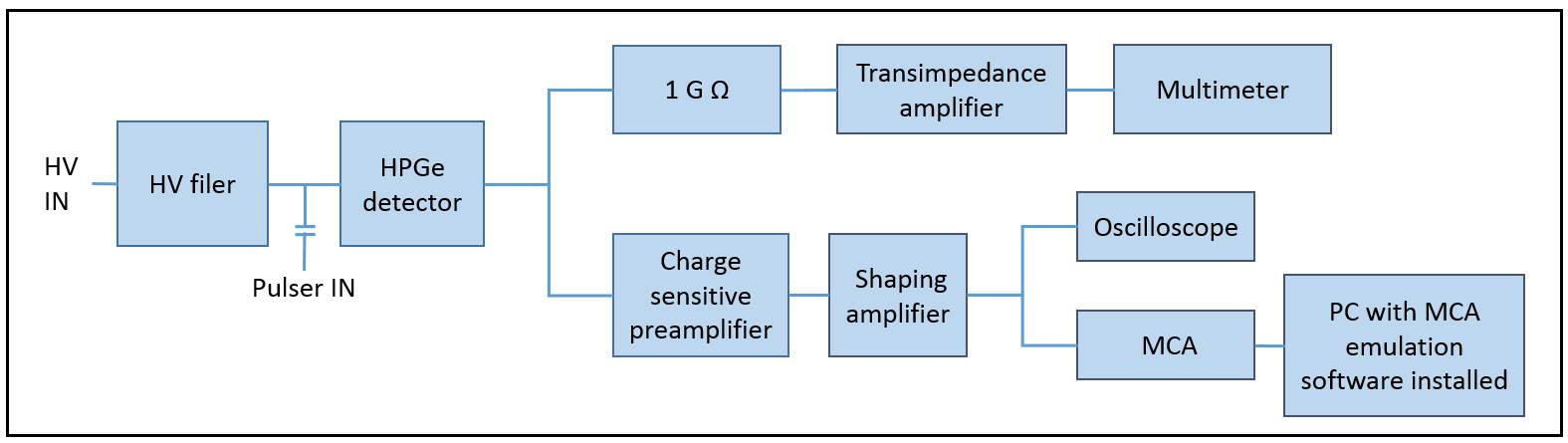}
  \caption{Diagram of how to conduct the electrical and spectroscopy measurements with the electronics shown in figure~\ref{fig:setup}. The detector was virtually grounded through the transimpedance amplifier.}
  \label{fig:dia}
\end{figure}

As an example, we chose one detector to discuss the performance in detail. The planar detector discussed in this article will be referred to as USD-L01 and its dimension is 1.86 cm (1.42 cm) in length and 1.24 cm (0.92 cm) in width for the bottom (top) surface. The thickness is 0.54 cm. After each detector was fabricated, it was loaded onto the variable-temperature sample stage in a test cryostat, pumped to a vacuum level of 10$^{-6}$ mbar and then cooled down to liquid nitrogen temperature in order to conduct the electrical and spectroscopy measurements. The cryostat shown in figure~\ref{fig:cryo} was borrowed from Lawrence Berkeley National Laboratory (LBNL). The design of the cryostat is such that the detector and variable temperature stage are enclosed by an infrared shield held at a temperature near 77 K. The temperature of the detector sample stage can be controlled from 79 K to greater than 200 K~\cite{looker} with the temperature controller shown in figure~\ref{fig:setup}. Also shown in figure~\ref{fig:setup} are the signal processing electronics and measurement electronics. The diagram presented in figure~\ref{fig:dia} shows how we did the electrical and spectroscopy measurements using the electronics shown in figure~\ref{fig:setup}. The detector was virtually grounded through the transimpedance amplifier. 

 To check the properties of USD-grown crystals and a-Ge contacts, both electrical measurements ($I$-$V$ and $C$-$V$ characteristics) and spectroscopy measurements are needed. 
 
 \section{Electrical Measurements}
 \label{sec:elec_mea}
 The following electrical measurements and calculations were conducted in this study: (1) leakage current versus applied voltage at 79 K, 90 K and 100 K; (2) the estimation of energy barrier heights for electrons and holes; (3) detector capacitance as a function of the applied voltage at 79 K; and (4) the determination of the detector impurity concentration. For all electrical measurements made in this work, a negative bias voltage was applied to the bottom contact of the detector USD-L01 and the signals were read out from the top electrical contact.
 
 \subsection{$I$-$V$ Characteristics}
 In the development of HPGe detectors, one important goal is to reduce the leakage current since it generates shot noise that degrades energy measurement. In general, there are three main contributions to the leakage current in a fully-depleted HPGe detector~\cite{looker}: (1) charge carrier injection at the electrical contact, i.e. hole injection at the positive contact and electron injection at the negative contact; (2) charge flow along detector side surfaces; and (3) thermal generation of electron-hole pairs in the detector. The contribution from the thermal generation can be reduced to a negligible level by cooling the detector to liquid nitrogen temperature. According to the study in ~\cite{hull,looker}, a guard ring structure can be used to separate the surface current component from the contact injections. No guard ring structure was fabricated for the detector USD-L01. Thus, the main contributions to the measured leakage current in this work are the first two sources.   
 
 As shown in figure~\ref{fig:dia}, the transimpedance amplifier was used together with a multimeter to measure the leakage current. There are two main functions of the transimpedance amplifier: (1) converting the current to voltage; (2) amplifying the voltage signal. The voltage read out from the multimeter can be converted back to the current according to the amplification factor selected by the transimpedance amplifier. In this work, the $I$-$V$ measurements were made for three temperatures, 79 K, 90 K and 100 K. The $I$-$V$ measurement at 79 K was used to check the performance of a-Ge contacts, while $I$-$V$ measurements at 90 K and 100 K allow us to estimate the electron and hole energy barrier of a-Ge contacts.
 
\begin{figure} [htbp]
  \centering
  \includegraphics[clip,width=0.9\linewidth]{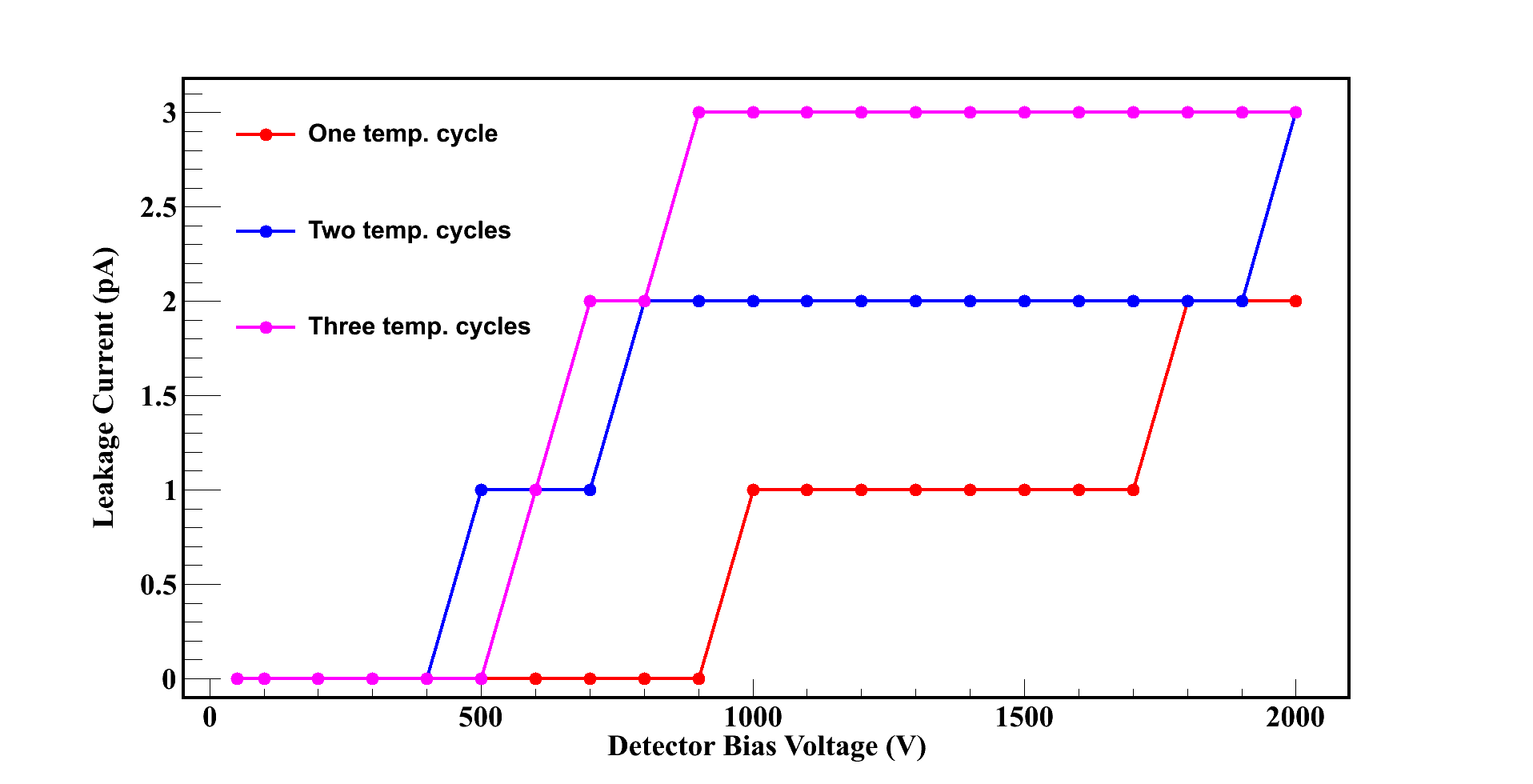}
  \caption{Measured leakage current as a function of bias voltage at 79 K after the detector USD-L01 went through one (red), two (blue), three (magenta) temperature cycles.}
  \label{fig:IV79K}
\end{figure}

\begin{figure} [htbp]
  \centering
  \includegraphics[clip,width=0.9\linewidth]{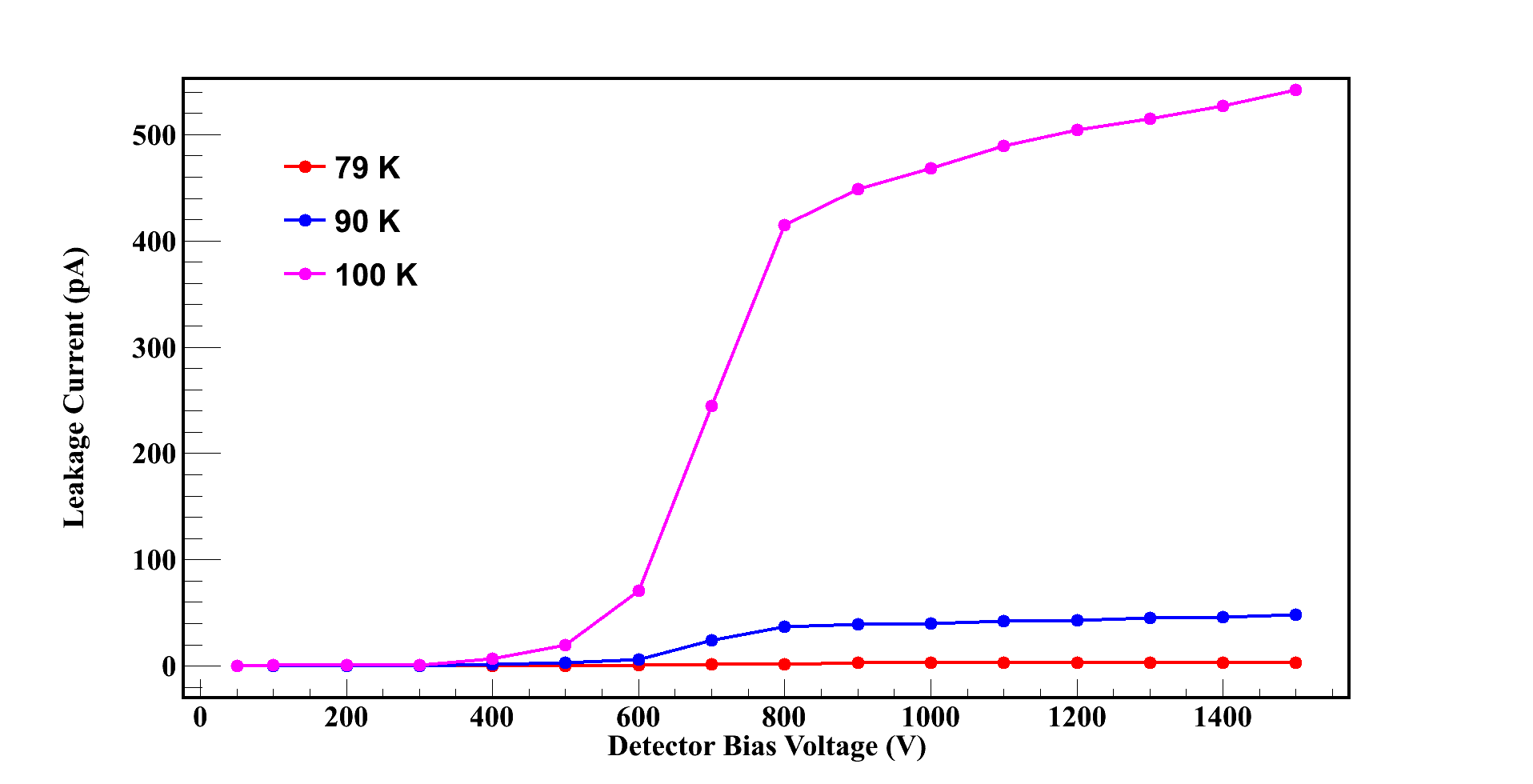}
  \caption{Measured leakage current as a function of bias voltage at 79 K, 90 K and 100 K after the detector USD-L01 went through three temperature cycles.}
  \label{fig:IV79_90_100K}
\end{figure}

 According to a series of studies conducted by Looker~\cite{looker} and Amman~\cite{amman5}, the following phenomenon about the leakage current stability for a-Ge contacts were observed: (1) the total time stored at room temperature rather than the temperature cycling is the relevant parameter affecting the change in the charge injection at the contacts; (2) the sputter pressure plays a key role in the leakage current stability for a-Ge contacts, and  a higher pressure produces a substantially more stable contact; (3) exposure to atmospheric air has little to no effect on the current change. Detector USD-L01 was made with the fabrication recipe recommended by Amman. Figure~\ref{fig:IV79K} presents the leakage current as a function of bias voltage after the detector USD-L01 went through one, two and three temperature cycles. The temperature cycle here is defined as warming from operating temperature to room temperature and then cooling back down to the desired temperature for testing. The wait period at room temperature between the first and third temperature cycle was about three months. As shown in figure~\ref{fig:IV79K}, there is little increase in the leakage current at 79 K for three months. This indicates that the a-Ge contacts have sufficient stability. Figure~\ref{fig:IV79_90_100K} displays the $I$-$V$ measurements at 79 K, 90 K and 100 K after the detector went through three temperature cycles. It is evident that the leakage current has strong dependence on the temperature and thus, cooling the Ge detectors down to liquid nitrogen temperature is an effective way of reducing leakage current. 

\subsubsection{Energy Barrier Heights of a-Ge layer}
One important physics parameter known as the energy barrier height is a key factor in determining the overall magnitude of the leakage current as a function of temperature~\cite{hull}. To understand the energy barrier heights of the a-Ge layer fabricated using USD equipment, we measured the leakage current of detector USD-L01 at various temperatures. 

A theoretical model about the current-voltage relationship for amorphous-crystalline semiconductor heterojunctions has been developed by D$\ddot{o}$hler and Brodsky during the 1970s~\cite{dohl,brod1,brod2}. This model has been successfully applied to a-Ge contacts on HPGe detectors~\cite{hull}. Based on the D$\ddot{o}$hler-Brodsky model and Schottky theory~\cite{scho,heni}, the leakage current density from electron injection at the bottom contact can be described as~\cite{looker}:
\begin{equation}
  J_c = J_0T^2exp[-(\varphi_e-\sqrt{\varepsilon_0\varepsilon_{Ge}/N_f}(V-V_{fd})/t)/kT],
  \label{e:j1}
\end{equation}
where the pre-factor $J_0$ is a constant to be determined from the measurements, $T$ is the temperature, $\varphi_e$ is the energy barrier height for electron, $k$ is the Boltzmann constant and 
\begin{equation}
\Delta\varphi = \sqrt{\varepsilon_0\varepsilon_{Ge}/N_f}(V-V_{fd})/t
\label{e:b1}
\end{equation}
is the barrier lowering term, with $\varepsilon_0 = 8.854\times10^{-14}F/cm$ the free-space permittivity, $\varepsilon_{Ge}$ = 16.2 the relative permittivity for Ge, $N_f$ the density of localized energy states (defects) near the Fermi level in a-Ge, $V$ the applied bias voltage, $V_{fd}$ the full depletion voltage and $t$ the detector thickness. The barrier lowering term $\Delta\varphi$ accounts for the lowering of the energy barrier height due to the electric field penetration into the a-Ge, since $\Delta\varphi$ increases as the applied voltage $V$ increases. To reduce the impact of the field penetration and keep a small value of $\Delta\varphi$ even at higher voltage, the a-Ge should be made with high density of localized energy states near the Fermi level. Therefore, the determination of $N_{f}$ can be accomplished through measuring $\Delta\varphi$.   

By letting
\begin{equation}
b=\sqrt{(\varepsilon_0\varepsilon_{Ge}/N_f)}/t
\label{e:b2}
\end{equation}
and then taking a natural log of both sides, eq.~\ref{e:j1} can be simplified as below:
\begin{equation}
  ln(J_c) = ln(J_0)+2ln(T)-\frac{\varphi_e}{kT}-\frac{bV_{fd}}{kT}+\frac{b}{kT}V,
  \label{e:j2}
\end{equation}

where, $J_c$ is in A/cm$^2$, $J_0$ is in A/cm$^2$/K$^2$, $T$ is in K, and the last three terms on the right hand side are all unitless. To figure out the two unknown parameters, $J_0$ and $\varphi_e$, the measurements of $J_c$-$V$ at two temperatures are required. In this study, we took the $I$-$V$ measurements at 90 K and 100 K and then extracted the data after full depletion, corresponding to the top of the current step in figure~\ref{fig:IV79_90_100K} for 90 K and 100 K. The $I$-$V$ measurements at 90K and 100 K, as shown in figure~\ref{fig:IV79_90_100K}, have a step feature distribution. The leakage current at the bottom of the step is very low since the detector is partially depleted and only the hole injection at the positive contact and surface charge flow contribute to the current. Once the depletion region reaches the bottom contact and the detector becomes fully depleted, the electron injection at the bottom contact becomes a main source to the leakage current and thus the leakage current increases quickly, which causes the step to rise. After the full depletion is reached, the leakage current continues to rise with increasing voltage, corresponding to the top of the step distribution. This can be attributed to the contributions from the contact injection due to the barrier lowering and the surface leakage increasing owing to a higher electric field. Thus, to find the energy barrier height for electrons, one needs to focus on the data at the top of the current step.

\begin{figure} [htbp]
  \centering
  \includegraphics[clip,width=0.9\linewidth]{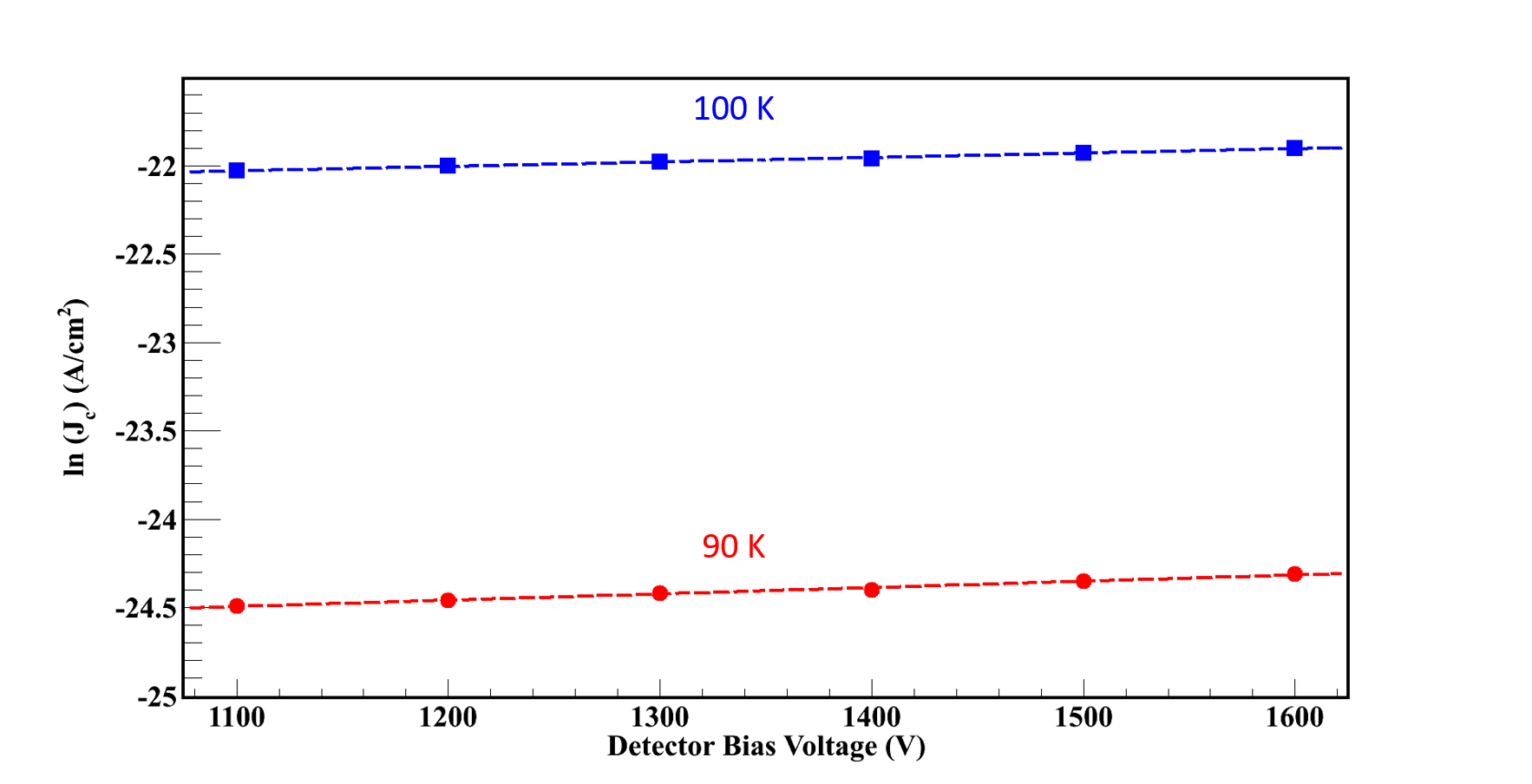}
  \caption{Measured leakage current as a function of bias voltage after full depletion at 90 K and 100 Kwith a linear fit to estimate the electron energy barrier.}
  \label{fig:fit}
\end{figure}

A linear fit was applied to the plotted data of $ln(J_c)$ as a function of the applied voltage $V$ as shown in figure~\ref{fig:fit}. The current density $J_c$ is related to the current $I$ through: $J_c=I/A$, with $A$ the active cross-section area of the detector.

The two constants in each linear fit function, $ln(J_c)$ = -24.89+0.0003571$V$ for 90 K and $ln(J_c)$ = -22.31+0.0002514$V$ for 100 K, correspond to the two terms, $ln(J_0)+2ln(T)-\frac{\varphi_e}{kT}-\frac{bV_{fd}}{kT}$ and $\frac{b}{kT}$ in eq.~\ref{e:j2}, respectively. Thus, we obtained: $\varphi_e$ = 0.18 eV and $J_0$ = 2.45$\times$10$^{-5}$ A/cm$^2$/K$^2$. From the fitted parameter, $p1$ = $b/kT$, one can obtain $b$ = 2.77$\times$10$^{-6}$ for $T$ = 90 K and $b$=2.16$\times$10$^{-6}$ for $T$ = 100 K. From eq.~\ref{e:b2}, we found that the value of $N_{f}$ is in a range of 4.0$\times$10$^{18}$ to 6.6$\times$10$^{18}$ eV$^{-1}$ cm$^{-3}$. The difference in the value of $N_{f}$ between 90 K and 100 K is partly due to the uncertainty from the fitting model since the determination of $\varphi_e$, $J_0$ and $N_{f}$ are correlated with each other from eq.~\ref{e:j2}. Another possible cause is a thin blade with a thickness of 0.5 mm used to cut the groove (the structure between the wing and the detector body).  For all the rest of our detectors, a blade with a thickness of 2 mm was used to cut the groove. This geometry difference would affect the sputtering of a-Ge on the side surfaces, which could result in a stronger temperature dependence of leakage current. Since only USD-L01 was made using a thin blade to cut the groove, more investigations are needed. 

\begin{figure} [htbp]
 \centering
  \includegraphics[clip,width=0.9\linewidth]{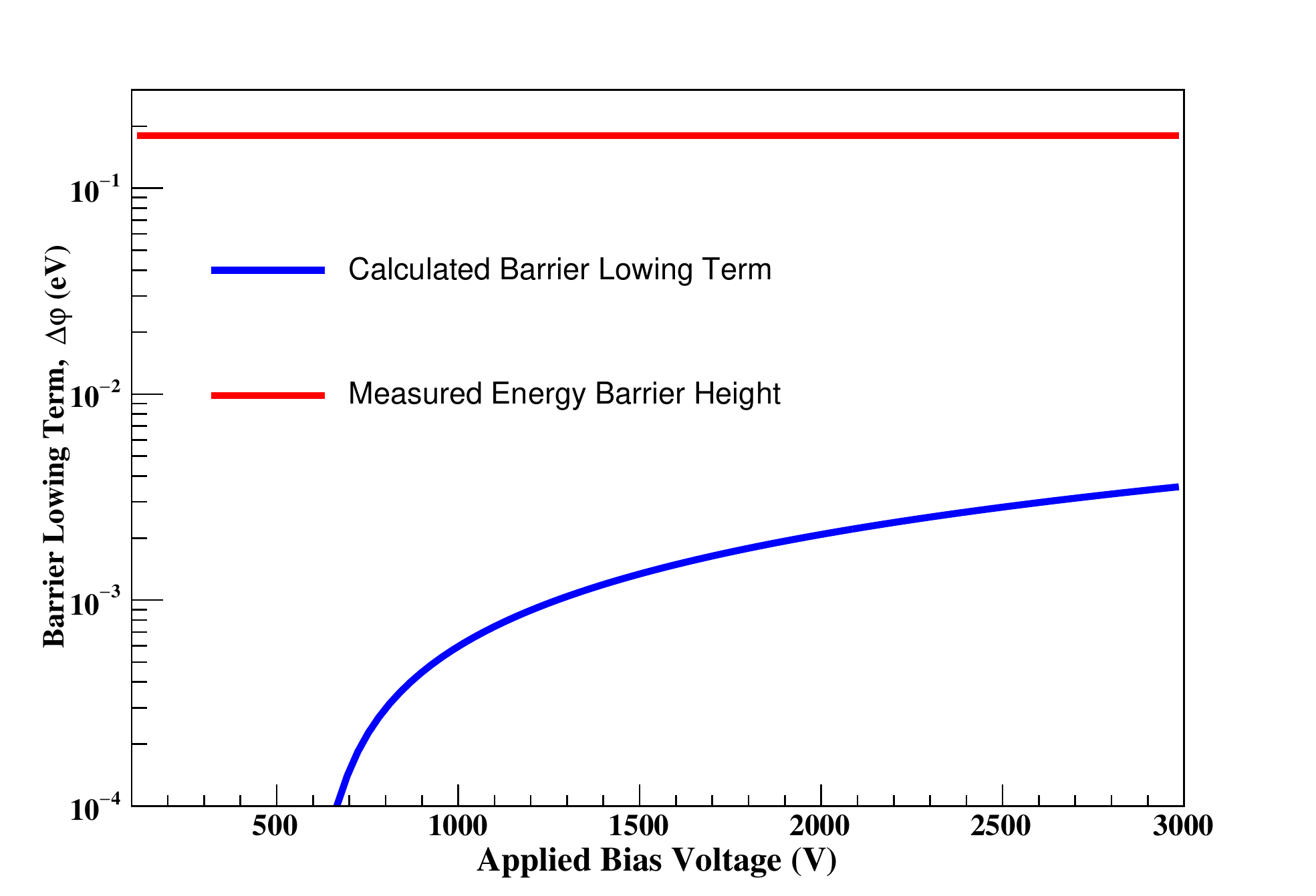}
  \caption{Calculated $\Delta\varphi$ as a function of applied voltage assuming $N_{f}$ = 4.0$\times$10$^{18}$ eV$^{-1}cm^{-3}$ for electron blocking. The measured energy barrier height is 0.18 eV.}
  \label{fig:barrierlowing}
\end{figure}

With $N_{f}$ determined, we were able to find out the relationship between the barrier lowering term ($\Delta\varphi$) and the applied bias voltage, which is shown in figure~\ref{fig:barrierlowing}. Also shown in figure~\ref{fig:barrierlowing} is the comparison between $\Delta\varphi$ and $\varphi_e$. Since the value of $\Delta\varphi$ is much less than 0.01 eV even at 3000 volts, the leakage current due to charge injection at the contacts mainly depends on $\varphi_e$ and $J_0$.

From simple Schottky contact theory, the sum of the energy barrier height of electron and hole ($\varphi_e$+$\varphi_h$) for a particular contact should equal the bandgap energy ($E_g$) of Ge. This has been verified by a previous study of amorphous semiconductor contacts on HPGe detectors~\cite{amman4}. In our case, $E_g$ = 0.73 eV at the average measurement temperature of 95 K. Therefore, the hole energy barrier of the contact is estimated to be, $\varphi_h$ = 0.55 eV, which is a factor of $\sim$2.7 higher than the electron energy barrier height. This explains why the contribution to the leakage current is mostly from the electron injection for our detector. However, due to a quite small prefactor ($J_0$), a small energy barrier height of electrons can still effectively block electron injection. 

Another detector, USD-L07, showed a value of $\varphi_e$ = 0.27 eV, which indicates the energy barrier height for holes is estimated to be about 0.48 eV. The variation of the energy barrier height for electrons and holes depends on the recipe for making a-Ge contacts as demonstrated by pioneers~\cite{amman1, amman2, looker}. Our results indicate that some parameters for making a-Ge contacts in the fabrication process at USD varied between the detector USD-L01 and USD-L07. This suggests that we need to continue to optimize our fabrication process in order to produce Ge detectors more uniformly in the future. 

 \subsection{$C$-$V$ Characteristics}
 The main purpose of the $C$-$V$ measurements (i.e. detector capacitance as a function of the detector bias voltage) is to determine the full depletion voltage of the detector and in turn the impurity concentration of the crystal. This can be done by applying a small voltage step from the pulser plus a constant voltage from the high voltage (HV) power supply, as shown in the diagram in figure~\ref{fig:dia}. The capacitance ($C$) of a planar Ge detector is similar to that of two flat, parallel metallic plates. That is:
 \begin{equation}
  C= \frac{\varepsilon_0\varepsilon_{Ge} A}{t_d},
  \label{e:c}
\end{equation}
where $A$ is the active cross-section area of the detector and $t_d$ is the depletion depth of the detector, which is related to the applied detector voltage ($V$) through:
\begin{equation}
  t_d=(\frac{2\varepsilon_0\varepsilon_{Ge}V}{qN})^{1/2},
  \label{e:d}
\end{equation}
where $q$ is the elementary charge and $N$ is the impurity concentration of the crystal. 

\begin{figure} [htbp]
  \centering
  \includegraphics[clip,width=0.9\linewidth]{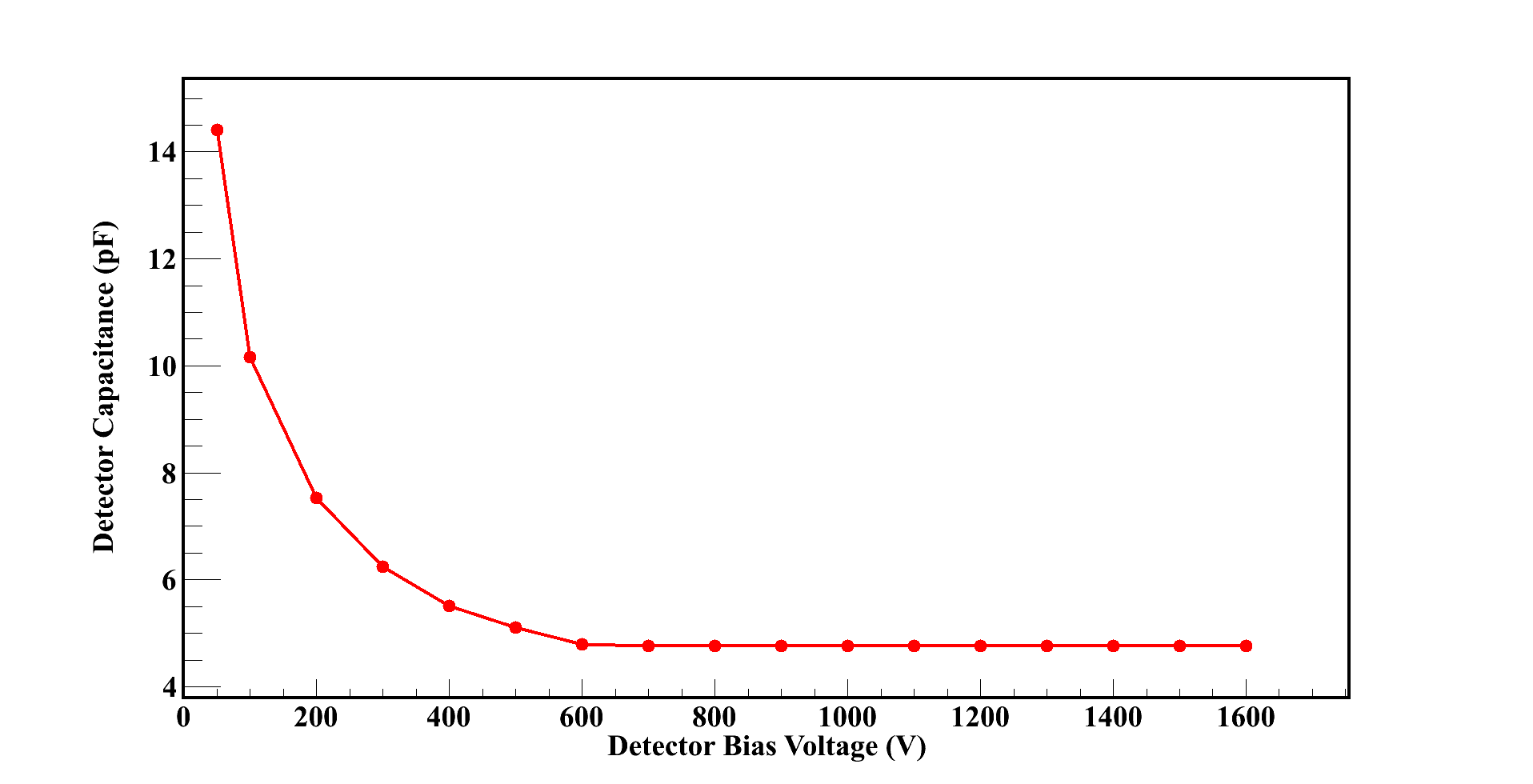}
  \caption{Measured detector capacitance as a function of bias voltage for detector USD-L01.}
  \label{fig:CV}
\end{figure}

As can be seen from eqs.~\ref{e:c} and ~\ref{e:d}, with the assumption of a constant impurity concentration, as the applied voltage ($V$) increases, the depletion depth ($t_d$) increases and the capacitance ($C$) between the top contact and the undepleted detector decreases until the detector is fully depleted. Once full depletion occurs, both $t_d$ and $C$ become constants with further increases in $V$ since $t_d$ and $C$ become detector thickness and capacitance, respectively. In this way, the full depletion voltage can be determined from the $C$-$V$ characteristics. As shown in figure~\ref{fig:CV}, the full depletion occurs at 700 V. For each applied voltage, a relative capacitance was determined from the oscilloscope by measuring the height of the pulse after the charge sensitive preamplifier and the shaping amplifier as presented in figure~\ref{fig:dia}. A negative bias voltage of 1600 V was then applied to the detector to measure its absolute capacitance through: $C=Q/V$, where $Q=Eq/\varepsilon$ with $E$ the energy of the peak generated from the pulser, $\varepsilon$ = 3 eV the average energy required to generate one electron-hole pair in Ge at 77 K, and $V$ the amplitude of the input pulse. The energy of the input pulse $E$ was determined from the energy calibration of a radiation source, Cs-137, which will be discussed in detail in the following section. In this way, the absolute detector capacitance is determined to be 4.76 pico-farads ($pF$). Then the relative capacitance at a given applied voltage was normalized to the absolute detector capacitance of 4.76 $pF$ at 1600 V, yielding the absolute detector capacitance for a given applied voltage as shown in figure~\ref{fig:CV}. According to eq.~\ref{e:c}, the theoretical value of the detector capacitance is about 3.43 $pF$, which is lower than the measured value (4.76 $pF$). This discrepancy is possibly due to the additional stray capacitance in the measurement, which was not measured in this work. Future work is needed to measure the stray capacitance, so that the detector capacitance can be determined more accurately.

\begin{figure} [htbp]
  \centering
  \includegraphics[clip,width=0.9\linewidth]{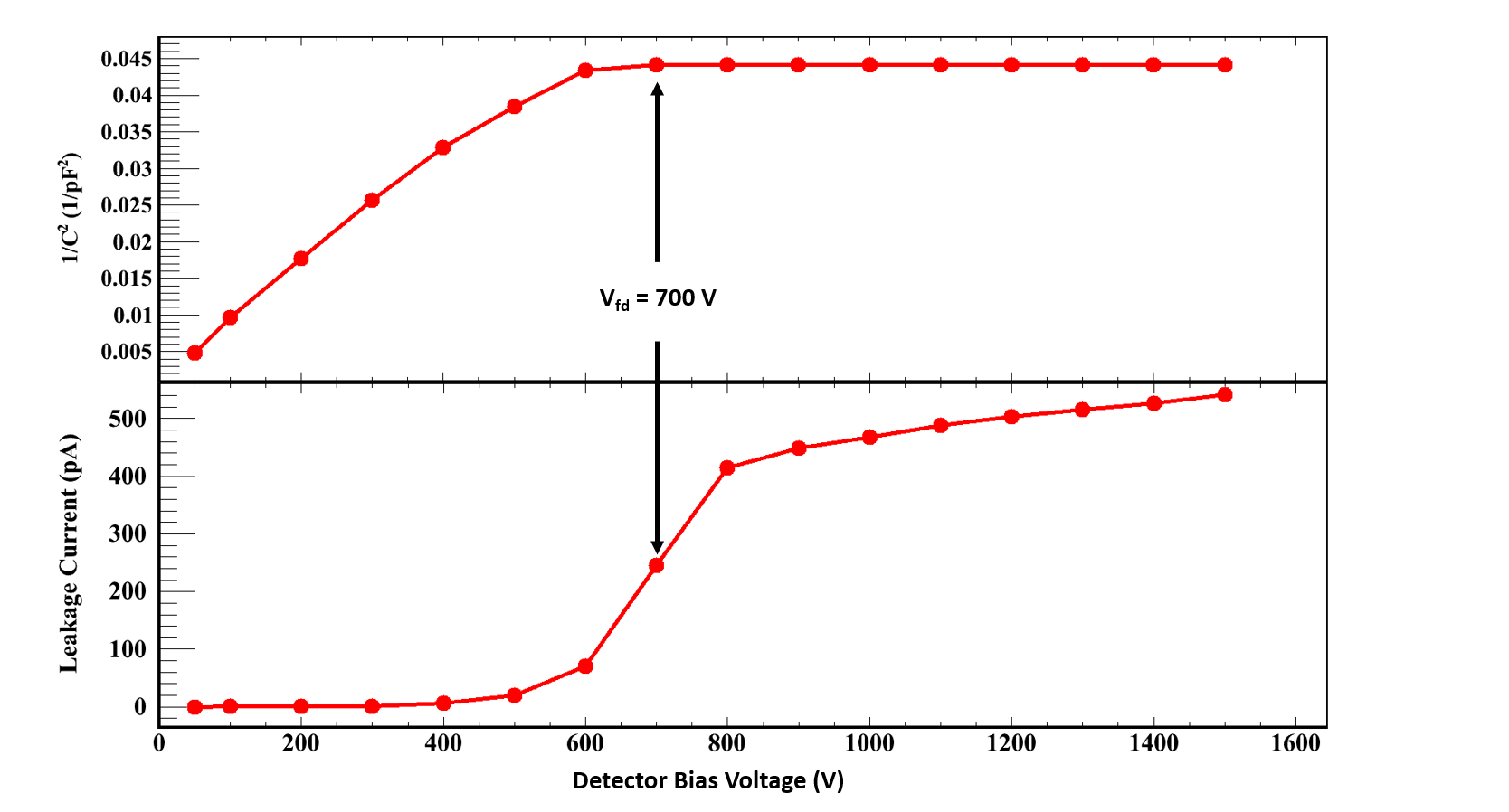}
  \caption{The determination of full depletion voltage for detector USD-L01 using $C$-$V$ measurements at 79 K (top) and $I$-$V$ measurements at 100 K (bottom).}
  \label{fig:IV_CV}
\end{figure}

As shown in figure~\ref{fig:IV_CV}, a step in the $I$-$V$ measurements at 100 K occurs when the detector is fully depleted as indicated by the 1/$C^2$-$V$ characteristic. This implies that, compared with surface leakage, the electron injection at the bottom contact is the main contribution to the measured leakage current. 

Once the full depletion voltage is determined, the impurity concentration of the detector can be estimated using eq.~\ref{e:d}, with $V$ = 700 V the full depletion voltage and $d$ =0.54 cm the detector thickness. The impurity concentration ($N$) of our detector USD-L01 was calculated to be $\sim$3.6$\times$10$^{10}$/cm$^{3}$, which is within a factor of 2 in agreement with the impurity concentration measured by the Hall effect measurement, (1.5$\sim$1.7)$\times$10$^{10}$/cm$^{3}$. Note that the impurity level measured by determining the full depletion voltage through $C$-$V$ measurements is more accurate than that of Hall Effect measurements, which use four ohmic contacts placed at the four corners of a Ge sample. Depending on the size of ohmic contacts and temperature of the measurements, there are errors associated with Hall effect measurements.

\section{Spectroscopy Measurements}
\label{sec:spec_mea}
\begin{figure} [htbp]
  \centering
  \includegraphics[clip,width=0.9\linewidth]{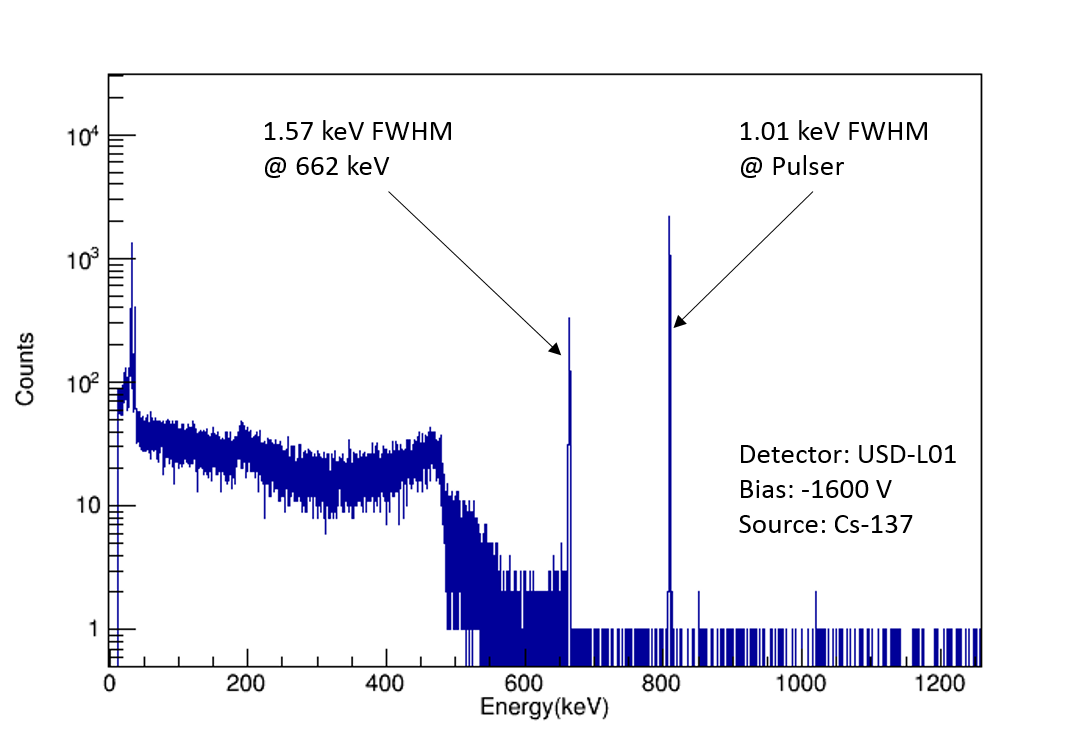}
  \caption{Energy spectrum from a Cs-137 source measured with the detector USD-L01. The source was positioned facing the detector bottom. The bias voltage of -1600 V was applied to the bottom electrical contact on the detector while the signals were measured from the top.}
  \label{fig:cs137}
\end{figure}

A radioactive source, Cs-137, with a radioactivity of 5.0 $\mu$Ci, was used to conduct the spectroscopy measurements. The energy spectrum of Cs-137 at 79 K was acquired with a negative bias voltage of -1600 V applied to the bottom electrical contact. The source was placed on the top of the cryostat, facing the top side of the detector. As shown in figure~\ref{fig:dia}, the signals were read out from the top electrical contact using a charge sensitive pre-amplifier operated at room temperature followed by an ORTEC 671 shaping amplifier set to an optimized shaping time of 3 $\mu$s. Figure~\ref{fig:cs137} shows the energy spectrum taken with the Cs-137 source in an hour. The x-axis is the energy in keV, which is converted from the ADC count using two photon peaks from Cs-137, 31.8~keV and 662~keV. Also shown in figure~\ref{fig:cs137} is an artificial peak due to the injected pulses from the high voltage line. The 662 keV peak was quite symmetrical since the applied voltage (1600 V) was much higher than the full depletion voltage (700 V) such that trapping effects of the charge carriers were minimized. The full width at half maximum (FWHM) of the 662 keV and the pulse peaks are 1.57 keV and 1.01 keV, respectively. The statistical driven energy resolution can be obtained using $\sqrt{1.57^2 - 1.01^2}$ = 1.2 keV at 662 keV, which is 0.18\%. This is a good energy resolution in comparison to commercial detectors~\cite{ortec, canberra}. Successful spectroscopy measurements and good energy resolution at 662 keV imply that the detector USD-L01 works very well. 
 
\section{Conclusion}
\label{sec:conl}
Driven primarily by the need of Ge detectors with  extremely low cosmogenic activation of Ge isotopes for ton-scale dark matter and neutrinoless double-beta decay experiments, USD has built the capacity in
zone refining~\cite{yang1, yang2, yang3}, crystal growth~\cite{wang1, wang2, wang3, wang4, wang5} and detector development to make Ge detector development in an underground laboratory possible. Using the USD-grown crystals, we have investigated the performance of the planar detectors fabricated with a-Ge contacts. Table~\ref{t:detectors} summarizes the number of the planar detectors we have fabricated at USD so far. These five detectors can all be fully depleted within 1200 volts, which indicate that the corresponding impurity level is below 4$\times$10$^{10}$/cm$^{3}$. Note that the detectors that cannot be fully depleted within 2000 Volts have higher impurity levels and they are not listed in Table~\ref{t:detectors}. The shown detectors have reasonable energy resolution at 662 keV. 
\begin{table}[htbp]
  \centering
  \caption{A summary of detector performance for all good planar Ge detectors made with USD-grown crystals at USD at liquid nitrogen temperature. $t_d$, $V_{fd}$ and $I_{fd}$ denote the detector thickness, full depletion voltage and leakage current at full depletion voltage, respectively. The active area is the area of the top contact surface. For detectors USD-L01, USD-L06, USD-L07 and USD-L08, the measured leakage current at full depletion voltage are mainly from the surface current and charge injection at both contacts since no guard ring was made for those detectors. For detector USD-W02, a guard ring was made on the top contact surface to separate the surface leakage from the charge injection. Hence, the measured leakage current from the center contact was solely from the charge injection at the contacts. The Hall effect measurements usually provide a range of the crystal impurity concentration, which is based on the measured impurity from two crystal samples (one is from the top and the other is from the bottom of the crystal). The FWHM at the peak from the pulser represents the noise level.}
\label{t:detectors}
  \setlength{\extrarowheight}{0.1cm}
  \begin{tabular}{|p{1.7cm}|p{0.6cm}|p{0.9cm}|p{0.7cm}|p{0.9cm}|p{2.0cm}|p{2.3cm}|p{1.2cm}|p{1.2cm}|} \hline
    Detectors&$t_d$ (cm) & Active Area (cm$^2$) &$V_{fd}$ (V) &$I_{fd}$ at 79 K (pA) &Crystal impurity concentration from $C$-$V$ measurements (/cm$^3$)&Crystal impurity concentration from Hall effect measurements (/cm$^3$)&FWHM at 662 keV (keV)&FWHM of pulser peak (keV)  \\ \hline 
    USD-L01&0.54 &1.31 &700 &1 &3.6$\times$10$^{10}$ &(1.5-1.7)$\times$10$^{10}$&1.57 &1.01  \\ \hline
    USD-L06&0.85 &1.72 &1200 &1 &2.9$\times$10$^{10}$&(2.6-4.0)$\times$10$^{10}$&2.22 &1.67   \\ \hline
    USD-L07&0.85 &2.00 &1000 &1 &2.2$\times$10$^{10}$ &(2.6-4.0)$\times$10$^{10}$&1.59 &1.19 \\ \hline
    USD-L08&0.85 &1.70 &800 &2 &1.9$\times$10$^{10}$  &(2.6-4.0)$\times$10$^{10}$&1.38 &1.03\\ \hline
    USD-W02&1.04&1.33 &500 &$<$1 &1.0$\times$10$^{10}$ &(0.5-3.0)$\times$10$^{10}$&2.97 &1.10  \\ \hline
  \end{tabular}
\end{table}

The characterization of those planar detectors made from USD crystals provides valuable feedback to the crystal growth process, such as the precise determination of the impurity level. The measured impurity levels using the Hall Effect for the USD-grown crystals agree with the impurity measurements reported in Table~\ref{t:detectors} using the $C$-$V$ and $I$-$V$ curves within a factor of 2. Currently, the detector fabrication at USD is limited to small detectors with a mass of $\sim$ 30 grams. However, with the characteristics of our crystals, larger detectors of $\sim$1 kg mass should be achievable. The largest detector-grade crystal grown at USD is about 2.2 kg at this stage. A large-size detector with a mass of greater than 3 kg would require an impurity level below 1.0$\times$10$^{10}$/cm$^3$. 

The detectors fabricated with a-Ge contacts allow us to study the electrical contact properties such as the barrier height for blocking electrons and holes as a function of the process parameters. Our results show that the detector fabricated at USD with USD-grown crystals have a higher energy barrier height for holes and a lower energy barrier height for electrons. However, due to the high density of localized energy states near the Fermi level in the a-Ge, the contacts we made can also block electron injection effectively at a high electric field. A detector contact with a guard-ring structure (USD-W02) demonstrates that the leakage current is dominated by the surface leakage current and the leakage current from the contact injection is negligible at liquid nitrogen temperature, as shown in Table~\ref{t:detectors}. The contacts have also gone through several thermal cycles, which have demonstrates the stability. We conclude that the good performance of the USD-made Ge detectors with a-Ge contact verify the USD crystal growing capability. This is a major milestone towards the development of Ge detectors with a-Ge contacts at USD for more scientific research in the future.

\acknowledgments
The authors would like to thank Dr. Mark Amman for his supervision on developing Ge planar detectors with a-Ge contacts at USD and Dr. Christina Keller for a careful reading of this manuscript. We would also like to thank Nuclear Science Division at Lawrence Berkeley National Laboratory for providing us a testing cryostat. This work was supported in part by NSF OISE 1743790, NSF OIA-1738695, NSF OIA-1738632, DOE grant DE-SC0004768 (DE-FG02-10ER46709), the South Dakota Board of Regents Innovation Grant, the Office of Research at the University of South Dakota and a research center supported by the State of South Dakota.
% We suggest to always provide author, title and journal data:
% in short all the information that clearly identify a document.
%\bibitem{a}
%Author, \emph{Title}, \emph{J. Abbrev.} {\bf vol} (year) pg.

%\bibitem{b}
%Author, \emph{Title},
%arxiv:1234.5678.

%\bibitem{c}
%Author, \emph{Title},
%Publisher (year).

% Please avoid comments such as "For a review'', "For some examples",
% "and references therein" or move them in the text. In general,
% please leave only references in the bibliography and move all
% accessory text in footnotes.

% Also, please have only one work for each \bibitem.

\end{document}